\documentstyle[preprint,aps]{revtex}
\begin{document}
\thispagestyle{empty}

\preprint{\vbox{
\hbox{TRI-PP-95-69}
\hbox{Oct 1995}
}}
\title{A note on Majorana neutrinos, leptonic CKM and\\
electron electric dipole moment}

\author{Daniel Ng and John N. Ng}
\address{
TRIUMF, 4004 Wesbrook Mall\\
Vancouver, B.C., V6T 2A3, Canada
}

\maketitle
\setlength{\baselineskip}{3ex}

\begin{abstract}
The electric dipole moment of the electron, $d_e$, is known to vanish up to
three-loops
in the standard model with massless neutrinos. However, if neutrinos are
massive Majorana particles, we obtain the
result that $d_e$ induced by leptonic CKM mechanism is non-vanishing at
two-loop order, and it applies to all massive Majorana neutrino models.
\end{abstract}

\vspace{3cm}

\newpage

\setlength{\baselineskip}{2.6ex}


The experimental searches for the electric diople moments of the
neutron, $d_n$, and the electron, $d_e$, have reached unprecedented
accuracies.  At present, their values \cite{1,2} are given by
\begin{equation}
d_n=(30\pm50)\times 10^{-27}{\rm \text{e-cm}} \; ,
\end{equation}
and
\begin{equation}
d_e=(1.8\pm 1.2 \pm 1.0 )\times 10^{-27} {\rm \text{e-cm}} \; .
\end{equation}
These values put important constraints on CP violating
theories beyond the standard model (SM).

Within the SM, CP violation is encoded in the complex elements of the
CKM unitary matrix, $V$, in the quark sector and observable effects are
proportional to \cite{3} the quantity
\begin{equation}
J = {\rm Im}(V_{ij}V^\ast_{kj}V_{kl}V^\ast_{il}) \; ,
\end{equation}
where the indices are not summed.  Obviously, $J$ is non-vanishing only
if not all the elements $V_{ij}$ can be made real and this implies
the existence of at least three generations of non-degenerate
massive quarks.
It is clear from Eq.(3) that
$J$ is antisymmetric about $i$ and $k$ as well as $j$ and $l$.
This leads to $J=0$ for degenerate quarks.
Using these properties,  Shabalin \cite{4}
showed that electric dipole moment of quarks
vanishes at two-loop level and hence implies a very small $d_n$.

Within the SM, the suppression of $d_e$ is even more severe than that of
$d_n$.  This is due in part to the masslessness of the three neutrinos and
hence there is no CKM type mixing in the lepton sector.  Any CP
violation effects for leptons will have to be induced from the quark
sector.  For the case of $d_e$, the lowest non-vanishing contribution
arises from the electric dipole moment of the $W$-boson, $d_W$.  The
authors of Ref. \cite{5} showed that $d_W$ vanishes at the two-loop
level and this implies that $d_e$ vanishes at the three-loop level.
Hence, the SM $d_e$ is estimated at the four-loop level to be \cite{6}
\begin{equation}
d_e \sim\frac{eG_F}{\pi^2} \left(\frac{\alpha}{2\pi}\right)^{3}m_{e}J
 \le 4\times 10^{-38} \; ,
\end{equation}
where J $\le 2\times 10^{-4}$ for the CKM elements is used. One factor of
$\alpha$ can be
replaced by $\alpha_s$, but this will not be a sufficient enhancement to be
experimentally
interesting.
We can see that it is
beyond experimental capabilities in the foreseeable future. At the same time if
offers the
opportunity that $d_e$ is a clean test of CP violation beyond the SM.

One can contemplate extending the SM by simply making the neutrinos
massive.  For instance, one can add a right-handed Dirac neutrino to each
family.  Then the neutrinos can have arbitrary but very small masses. Although
this is
unappealing, it is a possibility that is not ruled out.
Now the leptonic sector will exhibit the CKM type of mixing
as the analogue of the CKM matrix in the quark sector.
In complete parallel to the case of quarks,
$d_e$ can only be induced
at the three-loop level and is totally negligible since $d_e$ will be
proportional to the difference of masses squared of the neutrinos. For the
heaviest
allowed mass of  $30$ MeV for $\nu_\tau$, we anticipate a suppression factor of
$\frac{M\nu_{\tau}}{M^2_W}^2\le3\times10^{-7}$.

In this letter, we consider the situation when the neutrinos are massive
Majorana particles.  The simplest manifestation is the addition
of one or more \cite{7,8} right-handed neutrinos which then
allows us to construct
Majorana mass terms; however, for seesaw models of neutrino masses, more than
one
right-handed neutrino is required to generate a physical phase. Such Majorana
neutrinos are present in many grand unified
theories such as SO(10), E(6), etc.  They can also exist
in left-right symmetric models.  As opposed to the Dirac neutrinos where
there is only one CP violating phase for three generations
and other phases can be transformed
to the non-observable right-handed sector, there are many observable
phases associated with Majorana neutrinos. For example even with two
generations
of Majorana neutrinos there exists one CKM phase \cite{7}. For definiteness, we
consider $d_e$ as induced by a leptonic CKM mechanism which results from such
massive
Majorana neutrinos mixing.  Interestingly, this case has not been studied in
the literature and our main result is that $d_e$ is non-vanishing at the
two-loop level for Majorana neutrinos.

We begin our study by first noting the interaction lagrangian involving
the electric dipole moment is given by
\begin{equation}
{\cal L}=-\frac{1}{2}d_{e}F_{\mu\nu}\overline{e}i\sigma^{\mu\nu}\gamma_{5}e \;
,
\end{equation}
and the charged leptonic current interaction is
\begin{equation}
\textstyle{g\over \sqrt{2}}W^\mu U_{ij} \overline{\nu_i} \gamma_\mu
\textstyle{1-\gamma_5 \over 2} e_j + h.c. \; ,
\end{equation}
where $U_{ij}$ is the charged current mixing matrix analogue to the CKM
matrix in the quark sector. Here, the number of neutrinos are unrestricted.
The detailed structure of $U$ is
model-dependent but is not needed for our discussion.  What is of importance
is that the mixing matrix $U$ is unitary and  the
elements are in general complex. This is certainly true for the realistic case
of three
light neutrinos.  It is also obvious that the
one-loop contribution to $d_e$ is vanishing.

We now proceed to the two-loop calculation.  As was argued before, the
diagrams that are common to Dirac neutrinos need not be considered since
we know that they do not contribute to $d_e$; hence
we need only concentrate on those that
are specific to Majorana neutrinos.  The two-loop
$W$ exchange diagrams, depicted in Fig.~1a and 1b,
 give important contributions and illustrate the essential
physics at play here.
Notice that the diagrams do not exist in the SM
nor for massive Dirac neutrinos, since lepton number
conservation is not respected by the internal neutrino lines. An interesting
remark can
be made here. If one cuts across two $W$-bosons and the internal lepton line in
Fig.~1a,
and takes the latter to be the electron, one gets the process
$e^{-}e^{-}\rightarrow{W}^{-}W^{-}$
which is characteristic of Majorana neutrinos \cite{9,10}.
 After some standard manipulation, $d_e$ can be written in
terms of Feynman integrals given as
\begin{eqnarray}
\label{de_a}
d_e(a)&&= -{e~\alpha^2~m_e \over 256~\pi^2~s_W^4}~m_i~m_j~
J^l_{ij}~\int_0^1dx~\int_0^{1-x}dy~\int_0^1 ds~\int_0^{1-s} dt~
\int_0^{1-s-t} du \nonumber \\
&& { x(1-x)^2[(1-s)^2-(t+u)^2]+xy^2u(1-u)-x(1-x)y(1+3s+t+u-2tu-2u^2)
\over
\left[ m_i^2x(1-x)(1-s-t-u)+m_j^2(1-x-y)u+m_W^2(x(1-x)(s+t)+yu)+m_l^2xu
\right]^2 } \; ,
\end{eqnarray}
and
\begin{eqnarray}
\label{de_b}
d_e(b)&&= {e~\alpha^2~m_e \over 256~\pi^2~s_W^4}~m_i~m_j~
J^l_{ij}~\int_0^1dx~\int_0^{1-x}dy~\int_0^1 ds~\int_0^{1-s} dt~
\int_0^{1-s-t} du \nonumber \\
&& { x(1-x)^2[(1-s)^2-(t+u)^2]+xy^2u(1-u)-x(1-x)y(1+3s+t+u-2tu-2u^2)
\over
\left[ m_j^2x(1-x)(1-s-t-u)+m_i^2(1-x-y)u+m_W^2(x(1-x)(s+t)+yu)+m_l^2xu
\right]^2 } \; .
\end{eqnarray}
where the CP violating factor
$J^l_{ij}$ is given by
\begin{equation}
J^l_{ij}={\rm Im}(U^\ast_{je}U^\ast_{jl}U_{il}U_{ie})\ ,
\end{equation}
with $l$ being the internal charged lepton in the diagrams, where
$l=e,\mu,\tau$ for
the SM.  This is a variant
of the quark phase invariant factor. Note that $d_e(a)$ and $d_e(b)$ are
antisymmetric to each other
by interchanging $m_i$ and $m_j$, and this leads to nonvanishing EDM of an
electron.  It is important to notice some interesting properties
of $J^l_{ij}$.
$J^l_{ij}$ is antisymmetric about $i$ and $j$, namely
$J^l_{ij}=-J^l_{ji}$,
but symmetric about $e$ and $l$.  In addition, $J^l_{ii}=0$ which
implies at least two different massive Majorana neutrinos are needed
to generate two-loop non-vanishing $d_e$.

The situation is different in Fig.~1c.  The amplitude
is given as
\begin{eqnarray}
\label{de_c}
&&-e\left({g\over\sqrt{2}}\right)^4~J^l_{ij}~m_i~m_j~
\int{dk_1^4\over (2\pi)^4}~\int{dk_2^4\over (2\pi)^4}~
[4 (\not p_1 -\not k_1 -\not k_2)\gamma_\mu(\not p_2 -\not k_1 -\not
k_2) ]
{1\over k_1^2-m_i^2}{1\over k_2^2-m_j^2} \nonumber \\
&&{1\over (k_1-p_1)^2-m_W^2}
{1\over (k_2-p_2)^2-m_W^2}{1\over (p_2-k_1-k_2)^2-m_l^2}
{1\over (p_1-k_1-k_2)^2-m_l^2} \ .
\end{eqnarray}
We can see from Eq.(10) that aside from the factor $J^{l}_{ij}$, the intergrals
in
Eq.(10) are symmetrical about $m_i$ and $m_j$.
Thus Fig.~1c does not contribute to $d_e$ when the sum over $i$
and $j$ is taken.

In addition to the $W$-boson exchange diagrams, one has to include the
exchanges of would-be Goldstone-bosons.  These diagrams are multiplied by
factors of $(m_lm_i/m_W^2)^2 or (m_im_j/m_W^2)^2$ depending on whether one
or two Goldstone-bosons are exchanged.
For $m_i < m_W$, Goldstone-boson exchange diagrams are less important.
We can now give a semi-quantitative estimate of $d_e$ and
obtain
\begin{equation}
d_e \sim {\alpha^2~m_e m_i~m_j~(m_i^2-m_j^2) \over
256~\pi^2~s_W^4~m_W^6}~
J^l_{ij}~F(m_l^2/m_W^2,m_i^2/m_W^2,m_j^2/m_W^2)
(1.97 \times 10^{-16}) {\rm \text {e-cm}} \ ,
\end{equation}

\noindent
where all masses are taken in units of GeV.

The GIM factors $(m_i^2-m_j^2)/m_W^2$ for neutrinos are explicit and the
fact that $d_e$ vanishes when the neutrinos are massless is also
obvious.  Less obvious is the limit of degenerate charged lepton masses
where $d_e$ must also vanish.  For $m_l < m_W$, as the case in the SM,
this factor can be obtained by Taylor expanding the denominator in Eqs.(7,8)
 and a further factor of $m_l^2/m_W^2$ arises.  We can now give
numerical estimation for some cases.  If there exist Majorana neutrinos
with masses, $m_i^2 \le m^2_W$, and $|U_{ei}|^2 \le 10^{-2}$ obtained
from the charged current universality constraints \cite{8}, we
get $d_e \le 10^{-32}$ e--cm assuming that the integral function $F$ in
Eqs.(7,8) is of the order unity. In general the function F can only be
evaluated numerically; however, for the special limit of taken all masses
to be equal within the integral only it can be evaluated and we found it
to be 0.05.

Our calculation can be straightforwardly applied to $\mu$ and $\tau$.
There are enhancement factors in these cases.  For example, for
$d_\mu/d_e \sim m_\mu/m_e \sim 200$; and $d_\tau \sim 10^{-26}$ e--cm
without mixing suppression.
Unfortunately, the measurements of $d_\mu$ and $d_\tau$ are far more
difficult.

In conclusion, we find that in general Majorana neutrinos can induce
electric dipole moment at the two-loop level.  This is to be compared
to the Dirac neutrinos case in which contribution comes in at the
three-loop level.  It is a necessary condition to have two or more
non-degenerate Majorana neutrinos in order to contribute to $d_e$ at
this level.
However, the numerical estimates for $d_e \le 10^{-32}$ e--cm even
for Majorana neutrinos of masses in the 100 GeV range. This value is
several orders of magnitude lower than the current experiments can
reach in the near future. The suppression arises from two sources,
the first being the loop factor of $ \sim\frac{\alpha}{2\pi}$; the
second one arises from the mixing of heavy neutrinos with the light
ones which is constrained to be very small. This suppression factor $J^l_{ij}$
can be overcome if there exist heavy charged leptons that couple
to the heavy neutrinos with full strength. It is also less severe for the
$\tau$-lepton.
 This can enhance our result
by up to two orders of magnitude.

One of the authors, J.N. Ng, would like to thank Dr.W. Marciano for the kind
hospitality of the Institute of Nuclear Theory at the University of Washington
where this work is completed.
This work is supported in part by the Natural Science and Engineering Research
council of Canada.

\bibliographystyle{unsrt}

\begin{figure}
\caption{Two-$W$ exchange Feyman diagrams}
\label{fig:feydiag}
\end{figure}

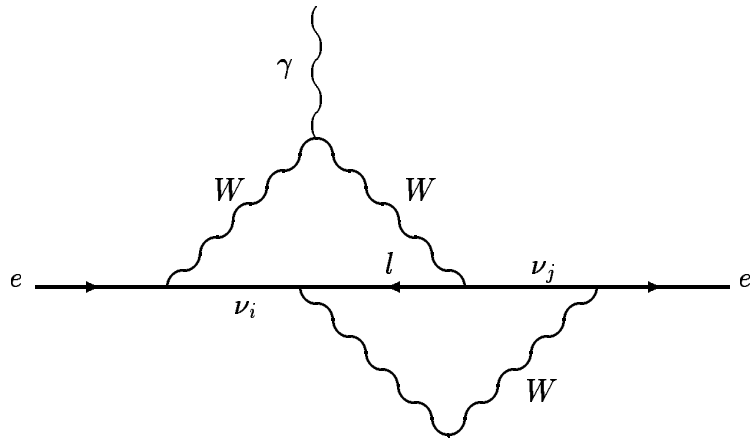
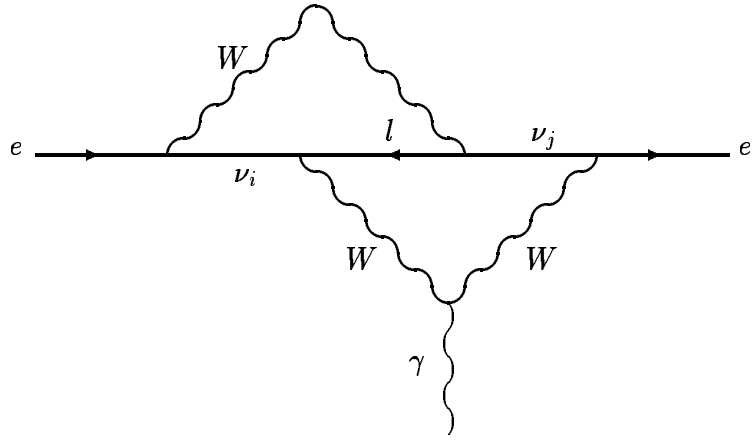
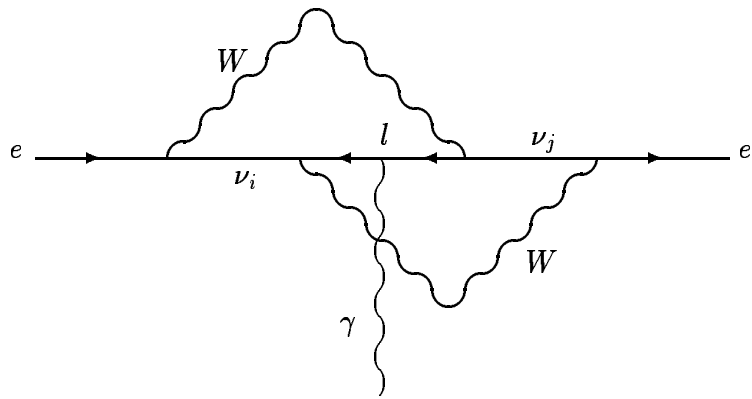
\end{document}